# Crystallization of Supercooled Liquid Elements Induced by Superclusters Containing Magic Atom Numbers

Robert F. Tournier,

CRETA /CNRS, Université Joseph Fourier, B.P. 166, 38042 Grenoble cedex 09, France. E-mail: robert.tournier@creta.cnrs.fr; Tel.: +33-608-716-878; Fax: +33-956-705-473.

**Abstract:** A few experiments have detected icosahedral superclusters in undercooled liquids. These superclusters survive above the crystal melting temperature $T_m$ because all their surface atoms have the same fusion heat as their core atoms and are melted by liquid homogeneous and heterogeneous nucleation in their core, depending on superheating time and temperature. They act as heterogeneous growth nuclei of crystallized phase at a temperature $T_c$ of the undercooled melt. They contribute to the critical barrier reduction, which becomes smaller than that of crystals containing the same atom number n. After strong superheating, the undercooling rate is still limited because the nucleation of 13-atom superclusters always reduces this barrier, and increases $T_c$ above a homogeneous nucleation temperature equal to $T_m/3$ in liquid elements. After weak superheating, the most stable superclusters containing n = 13, 55, 147, 309 and 561 atoms survive or melt and determine $T_c$ during undercooling, depending on n and sample volume. The experimental nucleation temperatures $T_c$ of 32 liquid elements and the supercluster melting temperatures are predicted with sample volumes varying by 18 orders of magnitude. The classical Gibbs free energy change is used, adding an enthalpy saving related to the Laplace pressure change associated with supercluster formation, which is quantified for n=13 and 55.

## 1. Introduction

An undercooled liquid develops special clusters that minimize the energy locally which are incompatible with space filling [1–3]. Such entities are homogeneously formed in glass-forming melts, and act as growth nuclei of crystals above the glass transition [4]. The formation of icosahedral nanoclusters has often been studied by molecular dynamics simulations into or out of liquids [5–8]. Silver superclusters containing the magic atom numbers n = 13, 55, 147, 309, 561 are more stable. Their formation temperature out of melt and their radius have been determined [5]. Icosahedral gold nanoclusters do not premelt below their bulk melting temperature [6]. Nanoclusters have been prepared out of liquids [9–14]. The density of states of conduction electrons at the Fermi energy being strongly reduced for particle diameters smaller than one nanometer leads to a gap opening [9,10]. Growth nuclei in melts are expected to have analogous electronic properties.

Superclusters containing magic atom numbers are tentatively viewed for the first time as being the main growth nuclei of crystallized phases in all liquid elements. I already considered that an energy saving resulting from the equalization of Fermi energies of nuclei and melts cannot be neglected in the classical crystal nucleation model [15,16]. An enthalpy saving $\varepsilon_v$ per volume unit of critical radius clusters equal to $\varepsilon_{ls} \times \Delta H_m/V_m$ was introduced in the Gibbs free energy change $\Delta G_{2ls}$ which gives rise to spherical clusters that transform the critical energy barrier into a less effective energy barrier, so inducing crystal growth around them at a temperature $T_c$ much higher than the theoretical homogeneous nucleation temperature equal to $T_m/3$. This enthalpy depends on $\Delta H_m$ the melting heat per mole at the melting temperature $T_m$, $V_m$ the molar volume and $\varepsilon_{ls}$ a numerical coefficient. The experimental growth temperature $T_c$ is often interpreted in the literature as a homogeneous nucleation temperature. This view is not correct because the $T_c$ of liquid elements is highly dependent on the sample volume v [17]. The crystallization temperatures are known to be driven by an effective critical energy barrier that is strongly weakened by the Gibbs free energy change associated with impurity clusters in the liquid [18,19]. The presence of $\varepsilon_v$ has for consequence to prevent the melting above $T_m$ of the smallest clusters acting as intrinsic growth nuclei reducing the critical energy barrier in undercooled liquids. The critical energy saving coefficient $\varepsilon_{ls}$ was shown for the first time as depending on $\theta^2 = [(T-T_m)/T_m]^2$ in liquid elements with a maximum at $T_m$ equal to 0.217 [15-16].

In this article, each cluster having a radius smaller than the critical radius has its own energy saving coefficient $\varepsilon_{nm}$ depending on $\theta^2$, n and its radius $R_{nm}$. In this case too, the cluster surface energy is a linear function of $\varepsilon_{nm}$ instead of a function of $\theta$ or T [20,21-25]. The Gibbs free energy change derivative $[d(\Delta G_{2ls})/dT]_p = -\Delta S_m$ at $T_m$ continues to be equal to the entropy change whatever the particle radius is because $(d\varepsilon_{ls}/dT)_{T=T_m}$ is equal to zero. All the surface atoms of growth nuclei have the same fusion heat as their core atoms [21]. They survive for a limited time above the melting temperature because they are not submitted to surface melting. A melt bath needs time to attain the thermodynamic equilibrium above the melting temperature $T_m$. This finding is the basic property permitting to assume for the first time that the growth nuclei in all liquid elements are superclusters instead of crystals. These superclusters are melted by homogeneous nucleation of liquid in their core instead of surface melting. A prediction of superheating effects is also presented for the first time for 38 liquid elements together with the predictions of undercooling rates depending on sample volumes and supercluster magic atom numbers n. The undercooling temperatures of gold and titanium have already been predicted using a continuous variation of growth nucleus radii and quantified values of $\varepsilon_v$ [22,23].

The equalization of Fermi energies of liquid and superclusters is not realized by a transfer of conduction electrons from nuclei to melts as I assumed in the past [15,16,24]. I recently suggested that a Laplace pressure change $\Delta p$ applied to conducting and nonconducting superclusters accompanied by an enthalpy saving per mole equal to $V_m \times \Delta p = \varepsilon_{ls} \times \Delta H_m$ is acting [25]. This quantity is proportional to $1/R_{nm}$ down to values of the radius $R_{nm}$, for which the potential energy is still equal to the quantified energy. Superclusters containing 13 and 55 atoms have an energy saving coefficient $\varepsilon_{nm0}$ which is quantified. This coefficient $\varepsilon_{nm0}$ associated with an n-atom supercluster strongly depends on n up to the critical number $n_c$ of atoms, giving rise to crystal spontaneous growth when $\varepsilon_{nm0}$ is equal to 0.217 in liquid elements [15].

The quantified values of $\varepsilon_v$ are known solutions of the Schrödinger equation which are obtained assuming that the same complementary Laplace pressure $\Delta p$ could be created by a virtual s-electron transfer from the crystal to the melt or from the melt to the crystal, creating a virtual surface charge screening associated with a spherical attractive potential [24]. All values of $\varepsilon_v$ for radii smaller than the critical values lead to a progressive reduction of electron s-state density as a function of n [23]. Reduced s-state density of superclusters depending on their radius and electronic specific heat of Cu, Ag and Au n-atom superclusters are studied, imposing a relative variation of Fermi energies during their formation in noble metal liquid state equal to $-2/3$ of the relative volume change. The radii of Ag superclusters calculated by molecular dynamics simulations in [5] are comparable with the critical radius values $R^*_{2ls}(T)$ deduced from this constraint.

This article follows the plan below:

2- The supercluster formation equations leading to crystallization
    2.1- Gibbs free energy change associated with growth nucleus formation,
    2.2- Thermal dependence of the energy saving coefficient $\varepsilon_{nm}$ of an n-atom condensed supercluster,
    2.3- Crystal homogeneous nucleation temperature and effective nucleation temperature.
3- The model of quantification of the energy saving of superclusters
4- Prediction of crystallization temperatures of 38 supercooled liquid elements at constant molar volume
5- Homogeneous nucleation of 13-atom superclusters and undercooling rate predictions
6- Maximum overheating temperature applied to melt superclusters at constant molar volume.
7- Electronic properties of Cu, Ag and Au superclusters
8- Silver supercluster formation into and out of undercooled liquid
9- Melting of Cu, Ag and Au superclusters, varying the overheating temperatures and times
    9.1- Superheating of Cu, Ag and Au superclusters
    9.2- Analysis of the influence of Cu superheating time on the undercooling rate
10- Conclusions.

## 2. Supercluster formation equations leading to crystallization

*2.1. Gibbs free energy change associated with growth nucleus formation*

The classical Gibbs free energy change for a growth nucleus formation in a melt is given in (1):

$$\Delta G_{1ls} = \theta \frac{\Delta H_m}{V_m} \frac{4\pi R^3}{3} + 4\pi R^2 \sigma_{1ls}, \quad (1)$$

where R is the nucleus radius, $\sigma_{1ls}$ the surface energy, $\Delta H_m$ the melting heat, $V_m$ the molar volume and $\theta = (T-T_m)/T_m$ the reduced temperature. Turnbull has defined a surface energy coefficient $\alpha_{1ls}$ in (2) which is equal to (3) [19,26]:

$$\sigma_{1ls}(\frac{V_m}{N_A})^{-1/3} = \alpha_{1ls} \frac{\Delta H_m}{V_m}, \quad (2)$$

$$\alpha_{1ls} = \left[\frac{N_A k_B \ln(K_{ls})}{36\pi \Delta S_m}\right]^{1/3}, \quad (3)$$

where $N_A$ is the Avogadro number, $k_B$ the Boltzmann constant, $\Delta S_m$ the melting entropy and $\ln(K_{ls}) = 90 \pm 2$.

An energy saving per volume unit $\varepsilon_{ls} \times \Delta H_m/V_m$ is introduced in (1); the new Gibbs free energy change is given by (4), where $\sigma_{2ls}$ is the new surface energy [15,27]:

$$\Delta G_{2ls} = (\theta - \varepsilon_{ls})\frac{\Delta H_m}{V_m}\frac{4\pi R^3}{3} + 4\pi R^2 \sigma_{2ls}. \quad (4)$$

The new surface energy coefficient $\alpha_{2ls}$ is given by (5):

$$\sigma_{2ls}(\frac{V_m}{N_A})^{-1/3} = \alpha_{2ls} \frac{\Delta H_m}{V_m}. \quad (5)$$

The critical radius $R^*_{2ls}$ in (6) and the critical thermally-activated energy barrier $\Delta G^*_{2ls}/k_B T$ in (7) are calculated assuming $(d\varepsilon_{ls}/dR)_{R=R^*_{2ls}} = 0$:

$$R^*_{2ls} = \frac{-2\alpha_{2ls}}{\theta - \varepsilon_{ls}}(\frac{V_m}{N_A})^{1/3}, \quad (6)$$

$$\frac{\Delta G^*_{2ls}}{k_B T} = \frac{16\pi \Delta S_m \alpha^3_{2ls}}{3N_A k_B (\theta - \varepsilon_{ls})^2 (1+\theta)} \quad . \tag{7}$$

They are not infinite at the melting temperature $T_m$ because $\varepsilon_{ls}$ is no longer equal to zero [15,16]. The homogeneous nucleation temperature $T_2$ (or $\theta_2$) occurs when the nucleation rate J in (8) is equal to 1, $\ln K_{ls}$ = 90 ± 2 in (9) and (10) respected with $\Delta G^*_{2ls}/k_B T$ = 90 neglecting the $\ln K_{ls}$ thermal variation [28]:

$$J = K_{ls} \exp(\frac{\Delta G^*_{2ls}}{k_B T}) \quad , \tag{8}$$

$$\frac{\Delta G^*_{2ls}}{k_B T} = \ln(K_{ls}) \quad . \tag{9}$$

The unknown surface energy coefficient $\alpha_{2ls}$ in (10) is deduced from (7) and (9):

$$\alpha^3_{2ls} = \frac{3N_A k_B (\theta_2 - \varepsilon_{ls})^2 (1+\theta_2) \ln(K_{ls})}{16\pi \Delta S_m} \quad . \tag{10}$$

The surface energy $\sigma_{2ls}$ in (5) has to be minimized to obtain the homogeneous nucleation temperature $T_2$ (or $\theta_2$) for a fixed value of $\varepsilon_{ls}$. The derivative $d\alpha_{2ls}/d\theta$ is equal to zero at the temperature $T_2$ (or $\theta_2$) given by (11), assuming that $\ln(K_{ls})$ does not depend on the temperature:

$$\theta_2 = \frac{T_2 - T_m}{T_m} = \frac{\varepsilon_{ls} - 2}{3} \quad . \tag{11}$$

The homogeneous nucleation temperature $T_2$ is equal to $T_m/3$ (or $\theta_2 = -2/3$) in liquid elements and $\varepsilon_{ls}(\theta)$ is equal to zero at this temperature [15,24].

The surface energy coefficient $\alpha_{2ls}$ is now given by (12), replacing $\theta$ by (11) in (10) for each value of $\varepsilon_{ls}$:

$$\alpha_{2ls} = (1+\varepsilon_{ls}) \left[\frac{N_A k_B \ln(K_{ls})}{36\pi \Delta S_m}\right]^{1/3} = (1+\varepsilon_{ls})\alpha_{1ls} \quad . \tag{12}$$

The classical crystal nucleation equation (4) is transformed into (13) with the introduction of the energy saving coefficient $\varepsilon_{ls}$:

$$\Delta G_{2ls}(R,\theta) = \frac{\Delta H_m}{V_m}(\theta - \varepsilon_{ls})4\pi \frac{R^3}{3} + 4\pi R^2 \frac{\Delta H_m}{V_m}(1+\varepsilon_{ls})(\frac{12k_B V_m \ln K_{ls}}{432\pi \times \Delta S_m})^{1/3} \quad . \tag{13}$$

The Laplace pressure p and the complementary Laplace pressure $\Delta p$ applied on the critical nucleus are calculated from the surface energy $\sigma_{2ls}$ with the equations (13) and (6) and $\Delta p$ is given by (14) [21,25]:

$$p = \frac{2\sigma_{2ls}}{R} = \frac{\Delta H_m}{V_m}[\theta - \varepsilon_{ls}(\theta)] ,$$

$$\Delta p = \frac{\Delta H_m}{V_m} \times \varepsilon_{ls}(\theta) = \frac{2(\delta\sigma_{2ls})}{R} \quad , \tag{14}$$

where $\delta\sigma_{2ls}$ is the complement proportional to $\varepsilon_{ls}$ in the surface energy in (13). The complement $\Delta p$ is equal to the energy saving $\varepsilon_{ls}(\theta) \times \Delta H_m/V_m$. The Gibbs free energy change $\Delta G_{2ls}$ in (13) directly depends on the cluster atom number n and the energy saving coefficient $\varepsilon_{nm}$ of the cluster instead of depending on its molar volume $V_m$ and its radius R as shown in (15):

$$\Delta G_{nm}(n,\theta,\varepsilon_{nm}) = \Delta H_m \frac{n}{N_A}(\theta - \varepsilon_{nm}) + \frac{(4\pi)^{1/3}}{N_A}\Delta H_m \alpha_{2ls}(3n)^{2/3} \quad . \tag{15}$$

The formation of superclusters having a weaker effective energy barrier than that of crystals precedes the formation of crystallized nuclei in an undercooled melt [5,29]. A spherical surface containing n atoms being minimized, a supercluster having a radius smaller than the critical radius cannot be easily transformed into a non-spherical crystal of n atoms because the surface energy would increase. The critical radius of superclusters could be larger than that of crystals because the supercluster density could be smaller, as already predicted for Ag [5] and confirmed in part 7. In these conditions, the transformation of a supercluster into a crystal is expected to occur above the critical radius for crystal growth when the Gibbs free energy change begins to decrease with the radius, while that of a supercluster increases up to its critical radius. It is shown in parts 3 and 4 that the supercluster energy saving $\varepsilon_{nm} \times \Delta H_m$ is quantified, depends on cluster radius R and atom number n, and is larger than the critical energy saving $0.217 \times \Delta H_m$. The cluster's previous formation during undercooling determines the spontaneous growth temperature $T_c$ reducing the effective critical energy barrier. The smallest homogeneously-condensed cluster controls the heterogeneous growth of crystals at temperatures higher than the homogeneous nucleation temperature $T_m/3$ ($\theta_2 = -2/3$) even in liquids which are at thermodynamic equilibrium at $T_m$ before cooling.

*2.2. Thermal dependence of the energy saving coefficient $\varepsilon_{nm}$ of an n-atom condensed cluster*

All growth nuclei which are formed in an undercooled melt are submitted to a complementary Laplace pressure. The energy saving coefficient $\varepsilon_{nm}$ of an n-atom cluster given in (16), being a function of $\theta^2$ as already shown [15], is maximum at $T_m$, with $(d\varepsilon_{nm}/dT)_{T=T_m}$ equal to zero:

$$\varepsilon_v = \varepsilon_{nm} \frac{\Delta H_m}{V_m} = \varepsilon_{nm0}(1 - \frac{\theta^2}{\theta_{0m}^2})\frac{\Delta H_m}{V_m}, \qquad (16)$$

where $\varepsilon_{nm0}$ is the quantified energy saving coefficient of an n-atom cluster at $T_m$ depending on the spherical nucleus radius R [24].

This thermal variation has for consequence that the fusion entropy per mole of a cluster of radius R is equal to the fusion entropy $\Delta S_m$ of the bulk solid [15,24]

$$\frac{3}{4\pi R^3}\left[\frac{d(\Delta G_{nm})}{dT}\right]_{T=Tm} = \frac{-\Delta S_m}{V_m}.$$

In these conditions, cluster surface atoms having the same fusion heat as core atoms, the cluster melts above $T_m$ by liquid droplet homogeneous nucleation above $T_m$ rather than by surface melting as expected for superclusters [6]. This $\theta^2$ thermal variation has already been used to predict the undercooling rate of some liquid elements [22,23].

The critical parameters for spontaneous supercluster growth are determined by an energy saving coefficient called $\varepsilon_{ls}$ in (17):

$$\varepsilon_v = \varepsilon_{ls}\frac{\Delta H_m}{V_m} = \varepsilon_{ls0}(1 - \frac{\theta^2}{\theta_{0m}^2})\frac{\Delta H_m}{V_m}, \qquad (17)$$

where $\varepsilon_{ls0} = 0.217$ is the critical value at $T_m$ and $\theta^{-2}_{0m} = 2.25$ in liquid elements [15,24]. A critical supercluster contains a critical number $n_c$ of atoms given by:

$$n_c = \frac{32\pi\alpha_{2ls}^3}{3(\theta - \varepsilon_{ls})^3}. \qquad (18)$$

*2.3. Crystal homogeneous nucleation temperature and effective nucleation temperature*

The thermally-activated critical energy barrier is now given by (19):

$$\frac{\Delta G^*_{2ls}}{k_B T} = \frac{12(1+\varepsilon_{ls})^3 \ln(K_{ls})}{81(\theta - \varepsilon_{ls})^2 (1+\theta)}, \quad (19)$$

where $\varepsilon_{ls}$ is given by (17). The coefficient of $\ln(K_{ls})$ in (19) becomes equal to 1 at the homogeneous nucleation temperature $T_m/3$ and the equation (9) and (11) are respected.

Homogeneously-condensed superclusters of n-atoms act as growth nuclei at a temperature generally higher than the homogeneous nucleation temperatures $T_m/3$ of liquid elements because they reduce the critical energy barrier as shown in (20) [18]:

$$\ln(J \times v \times t_{sn}) = \ln(K_{ls} \times v \times t_{sn}) - \left(\frac{\Delta G^*_{2ls}}{k_B T} - \frac{\Delta G_{nm}}{k_B T}\right) = 0, \quad (20)$$

where v is the sample volume, J the nucleation rate, $t_{sn}$ the steady-state nucleation time, $\ln K_{ls} = 90 \pm 2$, $\Delta G^*_{2ls}/k_B T$ defined in (19) and $\Delta G_{nm}$ in (15). The equation (20) is applied, assuming that n-atom superclusters preexist in melts when they have not been melted by superheating above $T_m$. It can also be applied when the homogeneous condensation time of an n-atom supercluster is evolved and its own critical energy barrier is crossed. The cluster thermally-activated critical energy barrier $\Delta G^*_{nm}/k_B T$ and the effective thermally-activated critical energy barrier $\Delta G_{neff}/k_B T$ of an n-atom supercluster are given by (21) and (22):

$$\frac{\Delta G^*_{nm}}{k_B T} = \frac{12(1+\varepsilon_{nm})^3 \ln(K_{ls})}{81(\theta - \varepsilon_{nm})^2 (1+\theta)}, \quad (21)$$

$$\frac{\Delta G_{neff}}{k_B T} = \frac{\Delta G^*_{nm}}{k_B T} - \frac{\Delta G_{nm}}{k_B T} = \frac{12(1+\varepsilon_{nm})^3 \ln(K_{ls})}{81(\theta - \varepsilon_{nm})^2 (1+\theta)} - \frac{\Delta G_{nm}}{k_B T}, \quad (22)$$

where $\Delta G_{nm}$ is given by (15), $\varepsilon_{nm}$ by (16) and $\ln K_{ls} = 90 \pm 2$. The quantified value $\varepsilon_{nm0} \times \Delta H_m$ of the cluster energy saving at $T_m$ is defined in the next part. The transient nucleation time being neglected, the growth around these nuclei is only possible when the steady-state nucleation time $t_{sn}$ is evolved and the relation (23) is respected:

$$\ln(J_n \times v \times t_{sn}) = \ln(K_{ls} \times v \times t_{sn}) - \frac{\Delta G_{neff}}{k_B T}, \quad (23)$$

where v is the sample volume and $t_{sn}$ the steady-state nucleation time. The crystallization follows this cluster formation time when, in addition, (20) is respected. The effective nucleation temperature deduced from (20) does not result from a homogeneous nucleation because it strongly depends on the sample volume v. This phenomenon explains why the effective nucleation temperature in liquid elements is observed around $\theta = -0.2$ in sample volumes of a few mm$^3$ instead of $\theta$ varying from $-0.58$ to $-0.3$ in much smaller samples [17,30].

## 3. Quantification of energy saving associated with supercluster formation

The potential energy saving per nucleus volume unit $\varepsilon_{ls} \times \Delta H_m/V_m$ is equal to the Laplace pressure change $\Delta p = 2 \times \delta \sigma_{ls}/R$ accompanying the transformation of a liquid droplet into a nucleus. The quantified energy is smaller than $2 \times \delta \sigma_{ls}/R$ at low radius R for n = 13 and 55. The calculation is made by creating a Laplace pressure on the surface of a spherical nucleus containing n atoms which would result from a virtual transfer of $n \times \Delta z$ electrons in s-states from the nucleus to the melt, $\Delta z$ being the fraction of transferred electrons per supercluster atom [31]. The potential energy $U_0$ would be equal to (24) and to zero beyond the nucleus radius R:

$$U_0 = -\frac{n \times \Delta z \times e^2}{8\pi\varepsilon_0 R}, \quad (24)$$

where e is the electron charge, and $\varepsilon_0$ the vacuum permittivity [32, p.135]. The quantified energy $E_q$ at $T_m$ is given by (25):

$$E_q = -\frac{n \times \varepsilon_{nm0} \times \Delta H_m}{N_A} \quad (25)$$

where $N_A$ is the Avogadro number. The quantified energy saving is given by (16) as a function of $\theta$ with $(\theta_{0m})^{-2}$ = 2.25 when $\varepsilon_{nm0}$ is known.

The Schrödinger equation depends only on the distance R of an s-state electron from the spherical potential center. The quantified solutions $E_q$ leading to the values of $\varepsilon_{nm0}$ are given by the two equations in (26), depending on $U_0$ which is equal to the complementary Laplace pressure $\Delta p$ acting on an n-atom cluster having a volume equal to $4\pi R^3/3$ through an intermediate parameter called k:

$$k = \frac{1}{\hbar} \times \left[2m(|U_0| - |E_q|)\right]^{1/2},$$
$$\frac{\sin(k \times R)}{k \times R} = \frac{\hbar}{(2 \times m_0 \times \varepsilon_0 \times R^2)^{1/2}}, \quad (26)$$

where $m_0$ is the electron rest mass and $\hbar$ Planck's constant divided by $2\pi$ [32, p.135].

The critical radii of liquid elements are sufficiently large at $T_m$ to assume that $U_0$ is equal to $E_q$ and to deduce the values of $\Delta z$ from the relation (24) = (25) with R = $R^*_{2ls}(\theta=0)$ and $\varepsilon_{nm0} = \varepsilon_{ls0} = 0.217$. The potential energy $U_0$ given by (24) is also equal to $-4\pi R^3/3 \times \Delta p$. Consequently, the $\Delta z$ in (24) does not depend on R at $T_m$. The value of $U_0$ is deduced from the atom number n which depends on molar volumes $V_m$ of solid elements extrapolated at $T_m$ from published tables of thermal expansion [16,33]. The values of $\varepsilon_{nm0}$ are calculated as a function of R using (27) instead of (26) for n ≥ 147 because $U_0$ is assumed to be equal to $E_q$:

$$\varepsilon_{nm0} = \frac{\varepsilon_{ls0} \times R^*_{2ls}}{R} \quad (27)$$

The condensed-cluster energy savings $\varepsilon_{nm0} \times \Delta H_m$ of 13 and 55 atoms are quantified and calculated from (25). The thermal variation of $\varepsilon_{nm}$ is given in (16) using these quantified values of $\varepsilon_{nm0}$.

4. **Prediction of crystallization temperatures $T_c$ of 38 undercooled liquid samples of various diameters**

The quantified and the potential energy saving coefficients $\varepsilon_{nm0}$ of silver clusters have been calculated using (26) and (27) and are represented in Figure 1 as a function of supercluster radius $R_{nm}$ which is assumed to continuously vary. These coefficients are equal for n ≥ 147. This last approximation is used in all liquid elements.

**Figure 1. The energy saving coefficient $\varepsilon_{nm0}$ versus the supercluster radius $R_{nm}$.** The quantified (square points) and non-quantified (diamond points) energy saving coefficients $\varepsilon_{nm0}$ are plotted versus the silver cluster radius. This coefficient is strongly weakened when R < 0.5 nm. Quantification is necessary for an atom number n < 147.

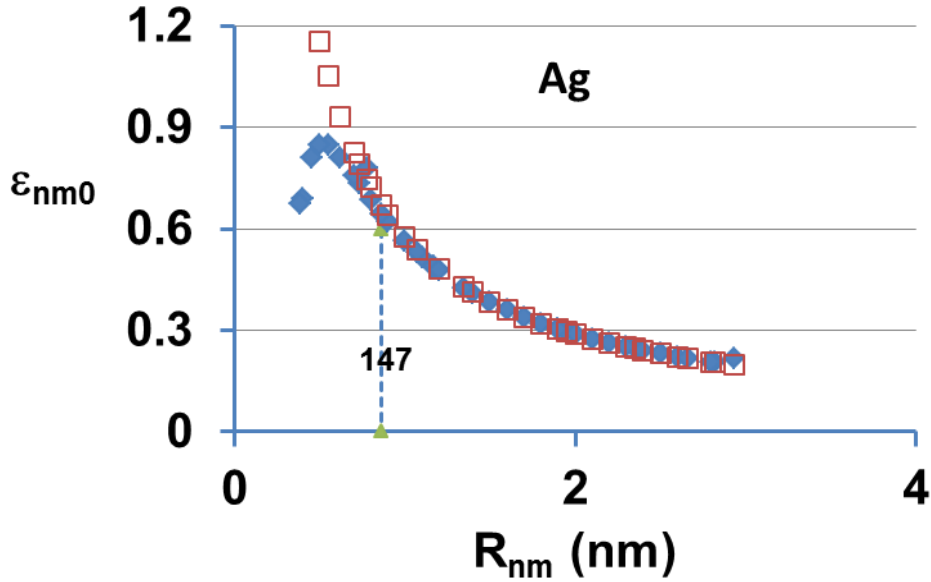

Properties of 38 elements are classified in Table 1:

*Column 1*- the liquid elements are classified as a function of their molar fusion entropy $\Delta S_m$,

*Column 2*- the molar volume of solid elements at $T_m$ in $m^3$,

*Column 3*- their fusion entropy $\Delta S_m$ in J/K/mole,

*Column 4*- their melting temperature $T_m$ in Kelvin,

*Column 5*- the atom magic number n of the supercluster inducing crystallization of the supercooled liquid at the closest temperature to the experimental crystallization temperature,

*Column 6*- the supercluster radius $R_{nm}$ in nanometers deduced from the molar volume $V_m$ using the relation (28):

$$n = \frac{4\pi R_{nm}^3}{3} \frac{N_A}{V_m}, \qquad (28)$$

*Column 7*- the energy saving coefficient $\varepsilon_{nm0}$ associated with the n-atom supercluster calculated using (28) for $n \geq 147$ and (26) for $n = 13$ and 55, with $\Delta z$ given in Table 2 column 3,

*Column 8*- the experimental reduced crystallization temperature $\theta_{c\,exp} = (T_c - T_m)/T_m$ of a liquid droplet having a diameter $D_{exp}$,

*Column 9*- the reduced crystallization temperature $\theta_{c\,calc}$ calculated using (20),

*Column 10*- the thermally-activated effective energy barrier $\Delta G_{eff}/k_B T$ given in (20) and (19) leading to the crystallization of the corresponding liquid element,

*Column 11*- the calculated diameter $D_{calc}$ in mm of the liquid droplet of volume v submitted to crystallization at $\theta_{c\,calc}$ using (20) and $v \times t_{sn} \cong v = \pi/6 \times D^3$ assuming that $t_{sn} = 1$ s,

*Column 12*- the experimental diameter $D_{exp}$ in millimeters of the liquid droplet crystallizing at $\theta_{cexp}$,

*Column 13*- references.

**Table 1. Reduced crystallization temperatures of 38 supercooled liquid elements** induced by condensed superclusters containing n = 13, 55, 147, 309 or 561 atoms.

| 1 | 2 | 3 | 4 | 5 | 6 | 7 | 8 | 9 | 10 | 11 | 12 | 13 |
|---|---|---|---|---|---|---|---|---|----|----|----|----|

| | $V_m \times 10^6$ | $\Delta S_m$ | $T_m$ | n | $R_{nm}$ | $\varepsilon_{nm0}$ | $\theta_c$ | $\theta_c$ | $\dfrac{\Delta G_{eff}}{k_B T}$ | D calc | D exp | Ref |
|---|---|---|---|---|---|---|---|---|---|---|---|---|
| | m$^3$ | J./K. | K | | nm | | Exp. | Calc. | | mm | mm | |
| Fe | 7.53 | 7.63 | 1809 | 55 | 0.55 | 0.859 | -0.304 | -0.298 | 61.7 | 0.10 | 0.10 | [34,35] |
| In | 15.90 | 7.69 | 429 | 147 | 0.98 | 0.707 | -0.26 | -0.266 | 51.0 | 0.0031 | 0.003 | [36,37] |
| Ti | 11.10 | 7.93 | 1943 | 309 | 1.11 | 0.546 | -0.18 | -0.191 | 70.6 | 1.93 | 1.80 | [38] |
| Zr | 14.60 | 7.95 | 2125 | 561 | 1.48 | 0.447 | -0.167 | -0.177 | 73.5 | 5.07 | 5.00 | [39,40] |
| Mn | 8.88 | 7.98 | 1517 | 309 | 1.03 | 0.545 | -0.206 | -0.217 | 59.8 | 0.05 | 0.05 | [30,41] |
| Pb | 18.80 | 8 | 600 | 147 | 1.03 | 0.698 | -0.26 | -0.249 | 57.0 | 0.02 | 0.02 | [17] |
| Co | 7.11 | 9.16 | 1768 | 55 | 0.54 | 0.815 | -0.27 | -0.28 | 63.7 | 0.19 | 0.20 | [36,42] |
| Ag | 11.00 | 9.16 | 1234 | 309 | 1.10 | 0.521 | -0.332 | -0.36 | 41.7 | 0.0001 | 0.0001 | [43] |
| Au | 10.80 | 9.43 | 1336 | 309 | 1.10 | 0.516 | -0.16 | -0.174 | 76.6 | 14.21 | 15.00 | [30,44] |
| Tc | 8.60 | 9.47 | 2430 | 55 | 0.57 | 0.843 | -0.24 | -0.252 | 79.3 | 5.64 | | [28] |
| Cr | 7.54 | 9.6 | 2176 | 309 | 0.97 | 0.512 | -0.13 | -0.18 | 73.5 | 5.02 | | [28] |
| Re | 9.50 | 9.62 | 3453 | 55 | 0.59 | 0.862 | -0.241 | -0.255 | 71.9 | 3.01 | 2.90 | [45] |
| Ir | 9.20 | 9.62 | 2716 | 309 | 1.04 | 0.512 | -0.19 | -0.183 | 72.1 | 3.16 | 3.30 | [28,46] |
| Mo | 10.00 | 9.63 | 2890 | 309 | 1.07 | 0.512 | -0.18 | -0.18 | 73.4 | 4.97 | 4.90 | [38,47] |
| Os | 8.85 | 9.64 | 3300 | 147 | 0.83 | 0.656 | -0.2 | -0.208 | 71.7 | 2.82 | | [28,48] |
| Pd | 9.91 | 9.64 | 1825 | 309 | 1.03 | 0.512 | -0.182 | -0.209 | 62.0 | 0.11 | 0.10 | [30,49] |
| Pt | 9.66 | 9.65 | 2042 | 309 | 1.06 | 0.512 | -0.185 | -0.184 | 71.6 | 2.69 | 2.60 | [38] |
| Cu | 7.57 | 9.66 | 1356 | 55 | 0.55 | 0.781 | -0.259 | -0.252 | 71.7 | 5.7 | 5.7 | [50-52] |
| Rh | 8.89 | 9.69 | 2239 | 147 | 0.80 | 0.654 | -0.204 | -0.209 | 71.1 | 2.30 | 2.30 | [38] |
| Ta | 12.40 | 9.74 | 3288 | 147 | 0.88 | 0.653 | -0.2 | -0.206 | 72.5 | 3.69 | 3.70 | [38,53] |
| Nb | 10.80 | 9.82 | 2740 | 309 | 1.10 | 0.509 | -0.176 | -0.179 | 73.7 | 5.42 | 5.00 | [38] |
| Hg | 14.20 | 9.91 | 232 | 13 | 0.42 | 0.000 | -0.38 | -0.549 | 51.6 | 0.0034 | 0.0035 | [17,43] |
| V | 8.93 | 10.07 | 2175 | 309 | 1.03 | 0.504 | -0.15 | -0.206 | 63.0 | 0.15 | 0.14 | [28] |
| Ni | 7.04 | 10.14 | 1726 | 55 | 0.54 | 0.791 | -0.278 | -0.276 | 62.7 | 0.14 | 0.14 | [34,36] |
| Ru | 8.75 | 10.19 | 2523 | 147 | 0.80 | 0.644 | -0.2 | -0.202 | 73.7 | 5.38 | 5.00 | [28] |
| Hf | 14.90 | 10.2 | 2500 | 309 | 1.22 | 0.502 | -0.18 | -0.179 | 73.8 | 4.8 | 4.60 | [38] |
| Gaβ | 13.40 | 10.31 | 256.2 | 13 | 0.39 | 0.000 | -0.5 | -0.528 | 58.6 | 0.035 | 0.036 | [17,54] |
| Cd | 9.51 | 10.44 | 594 | 309 | 1.18 | 0.498 | -0.19 | -0.228 | 57.1 | 0.021 | 0.020 | [17] |
| Zn | 10.60 | 10.53 | 693 | 309 | 1.05 | 0.497 | -0.19 | -0.19 | 68.6 | 0.981 | | [28] |
| Al | 10.20 | 11.48 | 932 | 309 | 1.09 | 0.483 | -0.19 | -0.236 | 57.0 | 0.021 | 0.020 | [17,55] |
| W | 16.50 | 12.69 | 3680 | 309 | 1.08 | 0.467 | -0.155 | -0.177 | 73.1 | 4.45 | 4.20 | [26,56] |
| Sn | 11.19 | 13.46 | 520 | 13 | 0.44 | 0.371 | -0.37 | -0.48 | 50.3 | 0.0022 | 0.0020 | [17,37] |
| Bi | 21.70 | 20.77 | 544 | 13 | 0.48 | 0.551 | -0.41 | -0.405 | 49.9 | 0.0019 | 0.0020 | [17,57] |
| Sb | 18.60 | 22.15 | 903 | 55 | 0.74 | 0.639 | -0.23 | -0.244 | 57.1 | 0.021 | 0.0200 | [17] |
| Te | 21.00 | 24.76 | 723 | 13 | 0.48 | 0.653 | -0.32 | -0.348 | 57.1 | 0.022 | 0.0200 | [17] |
| Se | 19.50 | 27.13 | 494 | 13 | 0.47 | 0.507 | -0.305 | -0.294 | 72.8 | 3.97 | 3.80 | [58] |
| Si | 12.20 | 29.79 | 1685 | 13 | 0.40 | 0.765 | -0.253 | -0.271 | 75.1 | 8.69 | 8.40 | [59-62] |
| Ge | 13.90 | 30.5 | 1210 | 13 | 0.415 | 0.695 | -0.39 | -0.387 | 41.0 | 0.0001 | 0.0001 | [36,43,52,63] |

The experimental reduced crystallization temperatures $\theta_{c\,exp}$ are plotted in Figure 2 versus the calculated $\theta_{c\,calc}$ using the supercluster atom-number n leading to about the same droplet diameter $D_{calc}$ as the experimental one $D_{exp}$. A good agreement is obtained between these values in 32 liquid elements in Figure 2 and Table 1. There is no good agreement for Hg, Sn, Al, Cd, V, and Cr because these elements are known to contain impurities or oxides. Their undercooling rates are too low compared to the calculated ones.

In Figure 3, the calculated droplet diameter logarithms are plotted as a function of those of experimental droplets used to study the undercooling rate. Six orders of magnitude are studied, corresponding to 18 orders of volume magnitude. Figure 3 shows that the model is able to describe the crystallization temperature dependence on the volume sample.

**Figure 2. Experimental undercooling temperatures versus calculated undercooling temperatures**. The experimental reduced crystallization temperatures $\theta_{c\,exp} = (T_c-T_m)/T_m$ are plotted versus the calculated ones $\theta_{c\ calc}$ of 38 liquid elements listed in Table 1. The smaller the atom number n, the smaller is the undercooling temperature, as shown in Table 1.

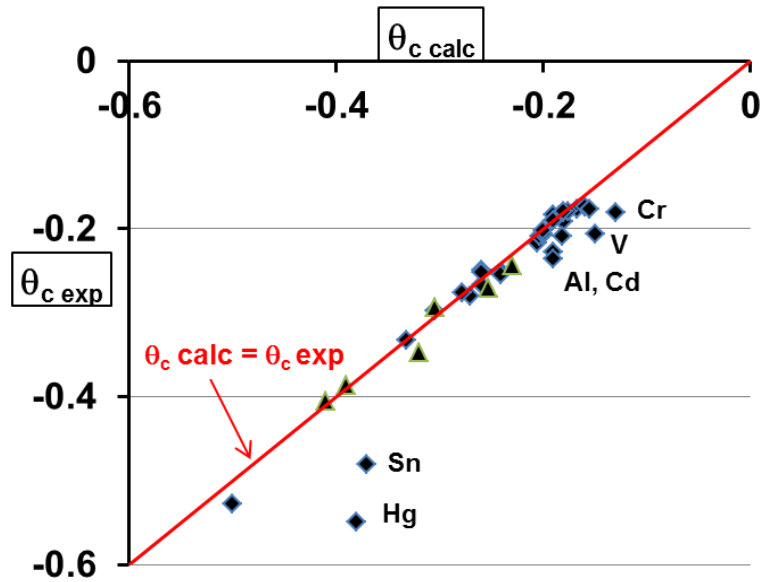

**Figure 3. Calculated liquid droplet diameters versus experimental liquid droplet diameters**. The calculated and experimental droplet diameters being crystallized are compared in a logarithmic scale. Smaller liquid droplets lead to lower undercooling temperatures, as shown in Table1.

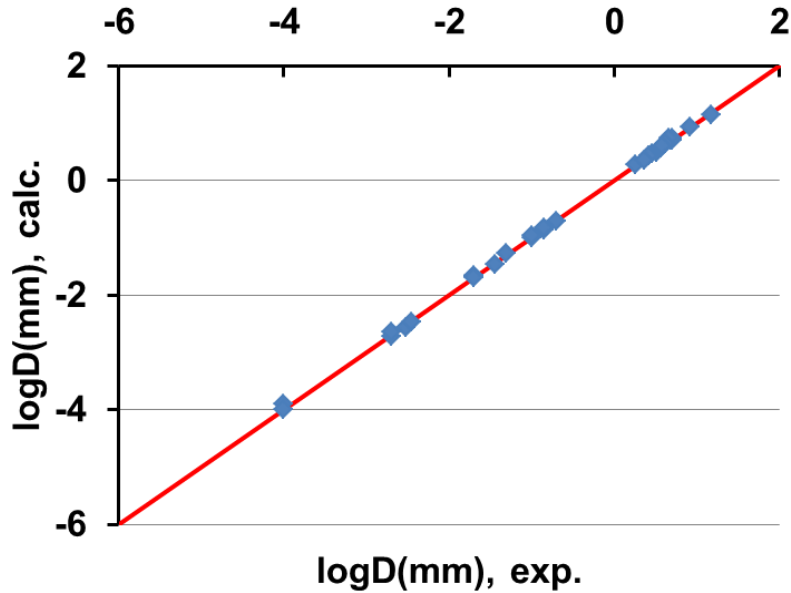

The atom numbers n of growth nuclei in liquid elements are represented in Figure 4 as a function of the reduced experimental crystallization temperature $\theta_{c\ exp}$ of liquid droplets having various diameters. The undercooling rate change is two times greater when the diameter varies from 0.036 to 8.4 mm and from 0.0001 to 5 mm for n = 13 and n = 309, respectively.

The experimental undercooling reduced temperature $\theta_c$ of gallium is the lowest of all the liquid elements and is equal to −0.58 and is a little higher than −2/3, corresponding to a crystallization temperature $T_c$ equal to 129K [17] and to a melting temperature of the α phase equal to 303 K. The gallium β phase is crystallized after undercooling. Its melting temperature is 257 K instead of 303 K for the α phase and its fusion entropy is 10.91

J/K/mole, as shown in Table 1, instead of 18.4 J/K/mole [54]. Its crystallization temperature of 129 K occurs in fact at $\theta_c = -0.5$. The calculated value is equal to the experimental one due to a previous condensation of 13-atom cluster which weakens the critical energy barrier. The model works without any adjustable parameter, and is able also to predict the nucleation rate of 13-atom clusters and the diameter of gallium droplets obtained with the liquid dispersion technique.

**Figure 4. N-atom superclusters acting as growth nuclei versus the experimental reduced undercooling temperatures.** The number n of atoms of superclusters is plotted versus the experimental reduced temperature of crystallization $\theta_{c\,exp}$.

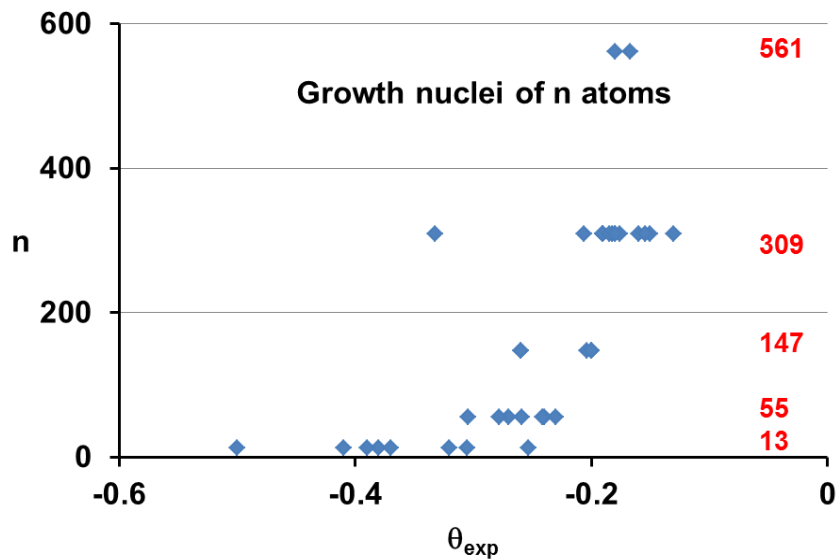

## 5. Homogeneous nucleation of 13-atom superclusters and undercooling rate predictions

Equations (20-23) are now used to calculate the homogeneous formation reduced temperature $\theta_{13c}$ of 13-atom clusters in a melt cooled below $T_m$ from thermodynamic equilibrium state at $T_m$ and the crystallization reduced temperature $\theta_c$ that they induce in liquid droplets of 10 micrometers in diameter. In Table 2, 38 liquid elements are considered. In 33 of them, the 13-atom cluster formation temperature is much larger than the crystallization temperature ($\theta_{13c} \gg \theta_c$). On the contrary, in indium, mercury, gallium β, cadmium and zinc, the two reduced temperatures are equal within the uncertainty on the energy saving coefficient value $\varepsilon_{13m0}$ given in Table 2. The crystallization temperatures of bismuth, selenium, tellurium, antimony, silicon and germanium with a growth around 13-atom clusters are predicted in good agreement with experimental values obtained with various sizes of droplets, as shown in Table 1.

In Figure 5, the homogeneous condensation reduced temperatures of 13-atom superclusters are compared with the reduced spontaneous growth temperatures which induce crystallization. The growth is organized around these 13-atom clusters which are formed, at temperatures higher than that of spontaneous crystallization. These homogeneous and heterogeneous crystallization temperatures depend on the droplet diameters. Their values given in Table 2, Column 10 are the lowest undercooling temperatures which can be obtained with 10 micrometer droplets.

**Figure 5. Condensation temperatures of 13-atom superclusters and crystallization temperatures in 10 micrometer droplets versus the melting entropy in J/K/mole**. The squares are the reduced formation temperatures of 13-atom superclusters which are ready to grow. These temperatures are larger than or equal to the reduced temperatures of spontaneous crystallization around them represented by diamond points.

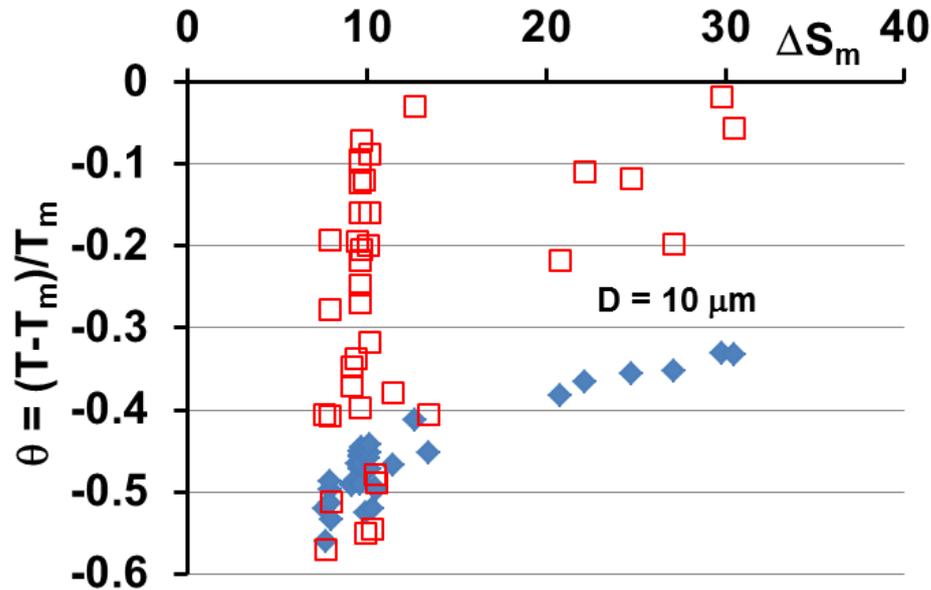

In Table 2, the liquid elements are still classified as a function of their fusion entropy $\Delta S_m$ given in Table 1 (Column 3):

*Column 1*- List of liquid elements,

*Column 2*- Critical radius for spontaneous growth,

*Column 3*- The number $\Delta z$ per atom of s-electrons virtually transferred from superclusters to melt at $T_m$,

*Column 4*- The energy saving coefficient $\varepsilon_{nm0}$ of 13-atom superclusters calculated with (24-26),

*Column 5*- The energy saving coefficient $\varepsilon_{nm0}$ of 55-atom superclusters calculated with (24-26),,

*Column 6*- The energy saving coefficient $\varepsilon_{nm0}$ of 147-atom superclusters calculated with (27) for n $\geq$ 147,

*Column 7*- The energy saving coefficient $\varepsilon_{nm0}$ of 309-atom superclusters,

*Column 8*- The energy saving coefficient $\varepsilon_{nm0}$ of 561-atom superclusters,

*Column 9*- The condensation reduced temperature of 13-atom superclusters in 10 micrometer droplets,

*Column 10*- The spontaneous growth reduced temperature around 13-atom superclusters in 10 micrometer droplets.

**Table 2. Energy saving coefficients $\varepsilon_{nm0}$ of n-atom superclusters,** and reduced condensation temperatures $\theta_{13c}$ of 13-atom superclusters inducing spontaneous growth at $\theta_c = (T_c-T_m)/T_m$ in 10 micrometer droplets.

| 1 | 2 | 3 | 4 | 5 | 6 | 7 | 8 | 9 | 10 |
| --- | --- | --- | --- | --- | --- | --- | --- | --- | --- |

| | $R^*_{2ls}$ | $\Delta z$ | $\varepsilon_{nm0}$ | $\varepsilon_{nm0}$ | $\varepsilon_{nm0}$ | $\varepsilon_{nm0}$ | $\varepsilon_{nm0}$ | $\theta_{13c}$ (10µm) | $\theta_c$ (10µm) |
|---|---|---|---|---|---|---|---|---|---|
| | nm | | n=13 | n=55 | n=147 | n=309 | n=561 | n=13 | n=13 |
| Fe | 2.48 | 0.107 | 0.67 | 0.859 | 0.709 | 0.553 | 0.454 | -0.406 | -0.52 |
| In | 3.18 | 0.033 | 0.09 | 0.707 | 0.707 | 0.552 | 0.452 | -0.57 | -0.56 |
| Ti | 2.79 | 0.134 | 0.86 | 0.881 | 0.700 | 0.546 | 0.448 | -0.277 | -0.498 |
| Zr | 3.05 | 0.161 | 0.99 | 0.900 | 0.699 | 0.546 | 0.447 | -0.192 | -0.487 |
| Mn | 2.58 | 0.098 | 0.64 | 0.842 | 0.698 | 0.545 | 0.447 | -0.408 | -0.514 |
| Pb | 3.32 | 0.050 | 0.37 | 0.786 | 0.698 | 0.545 | 0.446 | -0.513 | -0.534 |
| Co | 2.29 | 0.116 | 0.66 | 0.815 | 0.667 | 0.521 | 0.427 | -0.37 | -0.49 |
| Ag | 2.65 | 0.094 | 0.62 | 0.808 | 0.667 | 0.521 | 0.427 | -0.348 | -0.492 |
| Au | 2.61 | 0.103 | 0.67 | 0.808 | 0.660 | 0.516 | 0.423 | -0.337 | -0.485 |
| Tc | 2.41 | 0.174 | 0.89 | 0.843 | 0.659 | 0.515 | 0.422 | -0.195 | -0.466 |
| Cr | 2.30 | 0.150 | 0.80 | 0.826 | 0.656 | 0.512 | 0.420 | -0.248 | -0.47 |
| Re | 2.48 | 0.258 | 1.05 | 0.862 | 0.656 | 0.512 | 0.420 | -0.16 | -0.45 |
| Ir | 2.46 | 0.201 | 0.95 | 0.849 | 0.656 | 0.512 | 0.420 | -0.094 | -0.458 |
| Mo | 2.53 | 0.220 | 1.00 | 0.855 | 0.656 | 0.512 | 0.420 | -0.123 | -0.455 |
| Os | 2.42 | 0.250 | 0.76 | 0.819 | 0.656 | 0.512 | 0.420 | -0.269 | -0.45 |
| Pd | 2.52 | 0.133 | 1.04 | 0.861 | 0.656 | 0.512 | 0.420 | -0.098 | -0.472 |
| Pt | 2.49 | 0.154 | 0.85 | 0.832 | 0.655 | 0.512 | 0.419 | -0.217 | -0.467 |
| Cu | 2.30 | 0.094 | 0.54 | 0.781 | 0.655 | 0.511 | 0.419 | -0.397 | -0.49 |
| Rh | 2.42 | 0.164 | 0.86 | 0.834 | 0.654 | 0.511 | 0.419 | -0.204 | -0.465 |
| Ta | 2.70 | 0.266 | 1.09 | 0.864 | 0.653 | 0.510 | 0.418 | -0.071 | -0.445 |
| Nb | 2.57 | 0.216 | 1.00 | 0.850 | 0.652 | 0.509 | 0.417 | -0.119 | -0.451 |
| Hg | 2.81 | 0.020 | 0.00 | 0.525 | 0.650 | 0.507 | 0.416 | -0.551 | -0.526 |
| V | 2.40 | 0.164 | 0.85 | 0.823 | 0.646 | 0.504 | 0.413 | -0.2 | -0.459 |
| Ni | 2.21 | 0.121 | 0.66 | 0.791 | 0.645 | 0.503 | 0.413 | -0.318 | -0.472 |
| Ru | 2.37 | 0.190 | 0.91 | 0.829 | 0.644 | 0.502 | 0.412 | -0.16 | -0.453 |
| Hf | 2.83 | 0.225 | 1.04 | 0.846 | 0.643 | 0.502 | 0.412 | -0.087 | -0.442 |
| Gaβ | 2.71 | 0.021 | 0.00 | 0.509 | 0.641 | 0.500 | 0.410 | -0.546 | -0.52 |
| Cd | 2.41 | 0.052 | 0.30 | 0.710 | 0.638 | 0.498 | 0.409 | -0.479 | -0.496 |
| Zn | 2.43 | 0.055 | 0.26 | 0.698 | 0.637 | 0.497 | 0.407 | -0.489 | -0.497 |
| Al | 2.32 | 0.081 | 0.49 | 0.733 | 0.619 | 0.483 | 0.396 | -0.379 | -0.467 |
| W | 2.67 | 0.338 | 1.04 | 0.797 | 0.598 | 0.467 | 0.383 | -0.03 | -0.412 |
| Sn | 2.56 | 0.058 | 0.37 | 0.675 | 0.587 | 0.458 | 0.375 | -0.406 | -0.452 |
| Bi | 2.53 | 0.089 | 0.55 | 0.628 | 0.508 | 0.396 | 0.325 | -0.218 | -0.382 |
| Sb | 2.35 | 0.147 | 0.70 | 0.639 | 0.497 | 0.388 | 0.318 | -0.109 | -0.366 |
| Te | 2.36 | 0.132 | 0.65 | 0.613 | 0.479 | 0.374 | 0.306 | -0.118 | -0.356 |
| Se | 2.23 | 0.094 | 0.51 | 0.575 | 0.464 | 0.362 | 0.297 | -0.197 | -0.353 |
| Si | 1.85 | 0.290 | 0.77 | 0.597 | 0.450 | 0.351 | 0.288 | -0.018 | -0.331 |
| Ge | 1.92 | 0.208 | 0.695 | 0.584 | 0.447 | 0.349 | 0.286 | -0.056 | -0.332 |

## 6. Maximum superheating temperatures of superclusters at constant molar volume

*6.1. Superheating and melting of n-atom superclusters by liquid homogeneous nucleation*

N-atom superclusters survive above the melting temperature $T_m$ up to an superheating temperature which is time-dependent. They can be melted by liquid homogeneous nucleation in their core instead of surface melting. The Gibbs free energy change associated with their melting at a temperature $T > T_m$ is given by (29):

$$\Delta G_{nm}(n,\theta,\varepsilon_{nm}) = \Delta H_m \frac{n}{N_A}(-\theta - \varepsilon_{nm}) + \frac{(4\pi)^{1/3}}{N_A}\Delta H_m \alpha_{2ls}(3n)^{2/3}, \quad (29)$$

where the energy saving coefficient $\varepsilon_{nm}$ is given in (16) even for $\theta > 0$. The fusion enthalpy has changed sign as compared to (15) and the equalization of Fermi energies always still leads to an energy saving. An n-atom supercluster melts when (30) is respected:

$$\ln(K_{sl}.v_n.t_{sn}) = \frac{\Delta G_{nm}}{k_B T}, \quad (30)$$

where $v_n$ is the n-atom supercluster volume deduced from its radius given in Table 1, (28) and $t_{sn}$ is the superheating time at its own melting temperature because the supercluster radius is much smaller than its critical radius. The time $t_{sn}$ is chosen equal to 600 seconds and $\ln K_{sl}$ to 90.

*6.2. Overheating and melting of n-atom superclusters by liquid heterogeneous nucleation*

Melting temperatures of superclusters are reduced by previous melting of a 13-atom droplet in their core. These entities melt when (31) is respected for n =13:

$$\ln(K_{sl}.v_n.t_{sn}) = \frac{\Delta G_{nm}}{k_B T} - \frac{\Delta G_{13m}}{k_B T} \quad (31)$$

where the critical energy barrier $\Delta G^*_{nm}/k_B T$ no longer exists and is replaced by $\Delta G_{nm}/k_B T$, $\varepsilon_{nm}$ in (16), $\varepsilon_{nm0}$ in Table 2, $t_{sn}$ = 600 s and $\ln K_{ls}$ = 90. The critical barrier is not involved in (31) because the n-atom supercluster radius is much smaller than the critical radius for liquid growth and $\Delta G^*_{nm} \gg \Delta G_{nm}$.

*6.3. Prediction of melting temperatures of superclusters in 38 liquid elements by melt superheating above $T_m$*

The reduced melting temperatures $\theta = (T-T_m)/T_m$ of superclusters depending on their atom number n are given in several columns of Table 3 and in Figure 6. They are calculated assuming that the molar volume is constant, $t_{sn}$ = 600 s. and $\ln K_{sl}$ = 90. The liquid elements having fusion entropy $\Delta S_m$ larger than 20 J/K/mole have a melting temperature which is determined by liquid homogeneous nucleation because the 13-atom clusters melt at higher temperatures while those with $\Delta S_m < 20$ J/K/mole are submitted to chain-melting.

**Figure 6. The melting temperatures of superclusters containing 13, 55, 147, 309 and 561 atoms.**
These melting temperatures are given in columns 10, 11, 12 and 13 of Table 3 versus $\Delta S_m$.

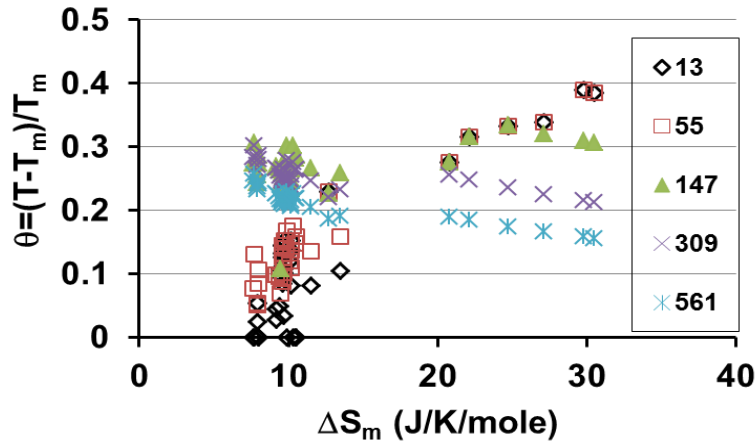

In Table 3,

*Column 1*- The liquid elements are classified as a function of the fusion entropy $\Delta S_m$.

*Column 2*- The melting temperature of 13-atom superclusters is high for large fusion entropies of Bi, Sb, Te, Se, Si and Ge. The highest value $\theta = 0.389$ is obtained for Si; the lowest $\theta = 0$ is obtained in 8 liquid elements. Homogeneous liquids do not contain any condensed cluster. They are crystallizing during undercooling because 13-atom clusters are condensed as shown in Figure 5. Their previous presence at $T_m$ does not have any influence on the undercooling rate.

*Column 3*- The calculated melting temperatures of 55-atom clusters induced by previous formation in their core of a droplet of 13 atoms are often lower than those of the 13-atom clusters. Then, they melt at the same temperature as the 13-atom clusters.

*Column 4*- The calculated melting temperatures of 147-atom clusters induced by previous formation in their core of a droplet of 13 atoms are larger than those of the 13-atom clusters from Fe to Al. They are nearly equal for W and smaller from Sn to Ge.

*Column 5*- The calculated melting temperatures of 309-atom clusters induced by previous formation in their core of a droplet of 13 atoms are larger those of the 13-atom clusters from Fe to Sn. They are smaller from Bi to Ge

*Column 6*- The calculated melting temperatures of 561-atom clusters induced by previous formation in their core of a droplet of 13 atoms are larger than those of the13-atom clusters from Fe to Sn except W. They are smaller from Bi to Ge.

*Columns 7, 8, 9 and 10*- The melting temperatures of 55-, 147-, 309- and 561-atom superclusters are obtained considering homogeneous liquid nucleation without introducing heterogeneous nucleation from 13-atom droplets.

*Columns 11, 12, 13 and 14*- The expected melting temperatures of 55-, 147-, 309- and 561-atom superclusters are selected in order to be coherent between them. The homogeneous nucleation temperature of some superclusters having a large fusion entropy are sometimes smaller than those of the 13-atom superclusters, as shown in Figure 6.

All these results have been obtained assuming that the superheating time at their own melting temperature is 600 seconds. The time effects on copper supercluster melting are examined in part 9 in relation with detailed experimental studies [51].

Table 3. The melting temperatures of superclusters.The final melting temperatures are given in Columns 2, 11, 12, 13 and 14. The temperatures in Columns 3, 4, 5 and 6 are calculated assuming that the melting starts from 13-atom droplets acting as heterogeneous nuclei in the core of superclusters. Those in Columns 7, 8, 9 and 10 correspond to a liquid homogeneous nucleation.

| 1 | 2 | 3 | 4 | 5 | 6 | 7 | 8 | 9 | 10 | 11 | 12 | 13 | 14 |
|---|---|---|---|---|---|---|---|---|---|---|---|---|---|
|   | $\theta_f$ | $\theta_f(n-13)$ n=55 | $\theta_f(n-13)$ n=147 | $\theta_f(n-13)$ n=309 | $\theta_f(n-13)$ n=561 | $\theta_f(n)$ Hom. n=55 | $\theta_f(n)$ Hom. n=147 | $\theta_f(n)$ Hom. n=309 | $\theta_f(n)$ Hom. n=561 | $\theta_f(n)$ n=55 | $\theta_f(n)$ n=147 | $\theta_f(n)$ n=309 | $\theta_f(n)$ n=561 |
|   | n=13 |   |   |   |   |   |   |   |   |   |   |   |   |
| Fe | 0 | 0.077 | 0.274 | 0.284 | 0.246 | 0.395 | 0.45 | 0.392 | 0.311 | 0.077 | 0.274 | 0.284 | 0.246 |
| In | 0 | 0.13 | 0.306 | 0.302 | 0.257 | 0.375 | 0.443 | 0.387 | 0.307 | 0.13 | 0.306 | 0.302 | 0.257 |
| Ti | 0.025 | 0.05 | 0.256 | 0.271 | 0.237 | 0.396 | 0.446 | 0.385 | 0.304 | 0.05 | 0.256 | 0.271 | 0.237 |
| Zr | 0.054 | 0.029 | 0.243 | 0.264 | 0.232 | 0.396 | 0.443 | 0.384 | 0.302 | 0.054 | 0.243 | 0.264 | 0.232 |
| Mn | 0 | 0.084 | 0.274 | 0.28 | 0.241 | 0.398 | 0.448 | 0.386 | 0.304 | 0.084 | 0.274 | 0.28 | 0.241 |
| Pb | 0 | 0.106 | 0.286 | 0.287 | 0.245 | 0.384 | 0.44 | 0.381 | 0.301 | 0.106 | 0.286 | 0.287 | 0.245 |
| Co | 0.045 | 0.099 | 0.27 | 0.267 | 0.227 | 0.42 | 0.446 | 0.37 | 0.286 | 0.099 | 0.27 | 0.267 | 0.227 |
| Ag | 0.027 | 0.098 | 0.268 | 0.266 | 0.226 | 0.412 | 0.442 | 0.367 | 0.285 | 0.098 | 0.268 | 0.266 | 0.226 |
| Au | 0.049 | 0.095 | 0.264 | 0.262 | 0.222 | 0.416 | 0.441 | 0.364 | 0.281 | 0.095 | 0.264 | 0.262 | 0.222 |
| Tc | 0.107 | 0.069 | 0.251 | 0.254 | 0.217 | 0.421 | 0.442 | 0.364 | 0.281 | 0.069 | 0.107 | 0.254 | 0.217 |
| Cr | 0.094 | 0.085 | 0.258 | 0.257 | 0.218 | 0.424 | 0.444 | 0.363 | 0.279 | 0.094 | 0.258 | 0.257 | 0.218 |
| Re | 0.145 | 0.047 | 0.238 | 0.246 | 0.212 | 0.422 | 0.442 | 0.362 | 0.278 | 0.145 | 0.238 | 0.246 | 0.212 |
| Ir | 0.126 | 0.061 | 0.246 | 0.25 | 0.215 | 0.422 | 0.442 | 0.362 | 0.279 | 0.126 | 0.246 | 0.25 | 0.215 |
| Mo | 0.133 | 0.054 | 0.242 | 0.248 | 0.213 | 0.422 | 0.44 | 0.361 | 0.278 | 0.133 | 0.242 | 0.248 | 0.213 |
| Os | 0.085 | 0.087 | 0.259 | 0.257 | 0.219 | 0.422 | 0.442 | 0.362 | 0.278 | 0.087 | 0.259 | 0.257 | 0.219 |
| Pd | 0.145 | 0.048 | 0.238 | 0.246 | 0.213 | 0.422 | 0.44 | 0.361 | 0.278 | 0.145 | 0.238 | 0.246 | 0.213 |
| Pt | 0.102 | 0.075 | 0.253 | 0.253 | 0.217 | 0.422 | 0.44 | 0.361 | 0.278 | 0.102 | 0.253 | 0.253 | 0.217 |
| Cu | 0.033 | 0.118 | 0.275 | 0.265 | 0.223 | 0.422 | 0.443 | 0.362 | 0.278 | 0.118 | 0.275 | 0.265 | 0.223 |
| Rh | 0.108 | 0.074 | 0.252 | 0.253 | 0.216 | 0.423 | 0.442 | 0.361 | 0.278 | 0.108 | 0.252 | 0.253 | 0.216 |
| Ta | 0.152 | 0.04 | 0.233 | 0.243 | 0.21 | 0.422 | 0.438 | 0.359 | 0.276 | 0.152 | 0.233 | 0.243 | 0.21 |
| Nb | 0.14 | 0.055 | 0.241 | 0.246 | 0.211 | 0.423 | 0.439 | 0.358 | 0.276 | 0.14 | 0.241 | 0.246 | 0.211 |
| Hg | 0 | 0.167 | 0.302 | 0.279 | 0.23 | 0.407 | 0.436 | 0.356 | 0.274 | 0.167 | 0.302 | 0.279 | 0.23 |
| V | 0.12 | 0.08 | 0.251 | 0.249 | 0.212 | 0.427 | 0.439 | 0.356 | 0.273 | 0.12 | 0.251 | 0.249 | 0.212 |
| Ni | 0.082 | 0.109 | 0.265 | 0.256 | 0.216 | 0.43 | 0.44 | 0.356 | 0.273 | 0.109 | 0.265 | 0.256 | 0.216 |
| Ru | 0.137 | 0.073 | 0.247 | 0.246 | 0.21 | 0.429 | 0.438 | 0.354 | 0.27 | 0.137 | 0.247 | 0.246 | 0.21 |
| Hf | 0.153 | 0.049 | 0.234 | 0.239 | 0.206 | 0.423 | 0.434 | 0.352 | 0.27 | 0.153 | 0.234 | 0.239 | 0.206 |
| Gaβ | 0 | 0.175 | 0.302 | 0.275 | 0.226 | 0.415 | 0.436 | 0.352 | 0.269 | 0.175 | 0.302 | 0.275 | 0.226 |
| Cd | 0 | 0.148 | 0.281 | 0.263 | 0.218 | 0.423 | 0.433 | 0.349 | 0.267 | 0.148 | 0.281 | 0.263 | 0.218 |
| Zn | 0 | 0.158 | 0.286 | 0.265 | 0.219 | 0.427 | 0.436 | 0.349 | 0.267 | 0.158 | 0.286 | 0.265 | 0.219 |
| Al | 0.082 | 0.136 | 0.266 | 0.247 | 0.205 | 0.435 | 0.428 | 0.337 | 0.256 | 0.136 | 0.266 | 0.247 | 0.205 |
| W | 0.23 | 0.072 | 0.227 | 0.22 | 0.186 | 0.445 | 0.42 | 0.323 | 0.243 | 0.23 | 0.227 | 0.22 | 0.186 |
| Sn | 0.105 | 0.158 | 0.259 | 0.233 | 0.191 | 0.443 | 0.411 | 0.314 | 0.205 | 0.158 | 0.259 | 0.233 | 0.191 |
| Bi | 0.275 | 0.151 | 0.214 | 0.185 | 0.151 | 0.455 | 0.358 | 0.256 | 0.19 | 0.275 | 0.275 | 0.256 | 0.19 |
| Sb | 0.315 | 0.136 | 0.203 | 0.177 | 0.145 | 0.454 | 0.35 | 0.248 | 0.184 | 0.315 | 0.315 | 0.248 | 0.184 |
| Te | 0.333 | 0.143 | 0.196 | 0.168 | 0.137 | 0.449 | 0.334 | 0.235 | 0.174 | 0.333 | 0.334 | 0.235 | 0.174 |
| Se | 0.339 | 0.16 | 0.193 | 0.163 | 0.133 | 0.448 | 0.321 | 0.225 | 0.166 | 0.339 | 0.321 | 0.225 | 0.166 |
| Si | 0.389 | 0.136 | 0.18 | 0.154 | 0.125 | 0.442 | 0.31 | 0.215 | 0.159 | 0.389 | 0.31 | 0.215 | 0.159 |
| Ge | 0.385 | 0.142 | 0.18 | 0.152 | 0.124 | 0.44 | 0.306 | 0.212 | 0.156 | 0.385 | 0.306 | 0.212 | 0.156 |

7. **Electronic properties of Cu, Ag and Au superclusters**

Electronic properties of superclusters can be calculated from the enthalpy saving associated with their formation temperature in noble metallic liquids because this energy is due to Fermi energy equalization of liquid and superclusters [23]. The Fermi energy difference $\Delta E_F$ between condensed superclusters of radius $R_{nm}$ containing n atoms and liquid state at $T_m$ can be directly evaluated for noble metals using (32) and assuming that $\Delta z$ is small, as shown in Table 2:

$$\frac{\Delta E_F}{2m} n\Delta z = \frac{n\varepsilon_{ls}\Delta H_m}{N_A}, \quad (32)$$

where m is the ratio of electron masses $m^*/m_0$, $m_0$ being the electron rest mass and $m^*$ the effective electron mass which is assumed to be the same in superclusters and liquid states, and $\Delta z$ being calculated at variable temperature using the known quantified energy saving $\varepsilon_{ls}$ in (17) and (25,26). The Fermi energy difference $\Delta E_F$ is plotted in Figure 7 as a function of $1/R^*_{2ls}$, where $R^*_{2ls}$ is given in (6) for Cu, Ag and Au assuming that the molar volume does not depend on temperature and a continuous variation of $R^*_{2ls}$. The quantified value $\varepsilon_{ls}$ is given in (17) and the $U_0$ and $\Delta z$ values are calculated with (25). For $R^* > 1$ nm, $\Delta E_F$ is proportional to the Laplace pressure, while for $R \ll 1$ nm there is a gap opening in the conduction electron band accompanying the quantification of the energy saving. This analysis is able to detect well-known properties of clusters out of the melt which become much less conducting at very low radii [10].

**Figure 7. Fermi energy difference $\Delta E_F$ between liquid and superclusters**. The $\Delta E_F$ in eV/mole is plotted as a function of the reverse of the critical radius $R^*_{2ls}$ in nm$^{-1}$.

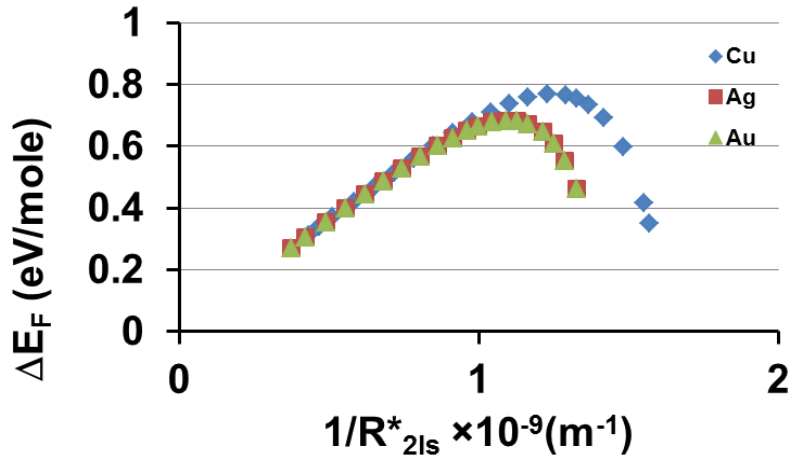

A strong variation of $\Delta E_F$ at constant molar volume $V_m$ is observed in Figure 7. In principle, the $\Delta E_F$ has to obey (33) in the liquid state because the Fermi energy $E_F$ depends on $(V_m)^{-2/3}$:

$$\frac{\Delta E_F}{E_F} = \frac{-2}{3}\frac{\Delta V_m}{V_m}, \quad (33)$$

where $E_F$ is the Fermi energy of the liquid, $V_m$ the molar volume of a supercluster of infinite radius, $\Delta V_m$ is the variation of the molar volume with the radius decrease. The supercluster molar volume $V_m$ has to depend on the particle radius instead of being constant. Equation (33) is respected when the formation temperature T of superclusters corresponding to the critical atom number $n_c$ in (18) and to a molar volume $V_m$ depending on $R^*_{2ls}$ is introduced. The formation temperatures of superclusters with magic atom numbers are indicated in Figure 8 using a special molar volume thermal variation $V_m(T)$ given in Figure 9 for each liquid element.

**Figure 8. The formation temperatures of superclusters containing n atoms**. The formation temperatures of Ag critical superclusters are plotted versus the critical number $n_c$ of atoms that they contain. The superclusters with magic atom numbers are represented by squares.

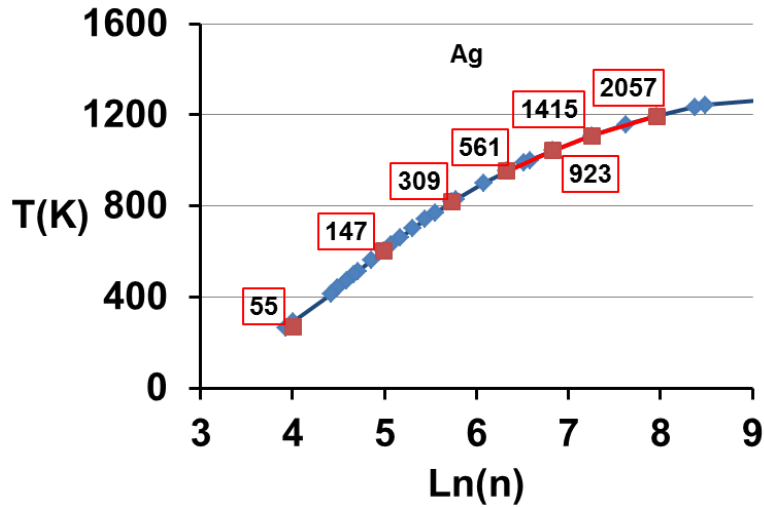

**Figure 9. The supercluster molar volume change of Cu, Ag and Au**. The supercluster molar volume change with the critical radius $R^*_{2ls}$ (being a hidden variable) is plotted as a function of their formation temperature in Kelvin up to $T_m$. Each point corresponds to a supercluster of radius $R^*_{2ls}$ and to an n-atom number.

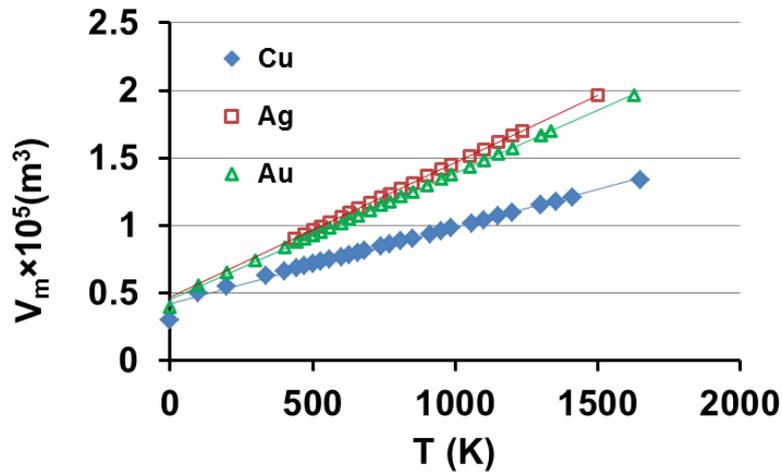

The following laws are used in Figure 8 and Figure 9:

$$V_m(m^3) = 7.37 \times 10^{-6} \times \theta + 11.8 \times 10^{-6} \text{ for Cu,} \qquad V_m(m^3) = 12.35 \times 10^{-6} \times \theta + 17 \times 10^{-6}$$

for Ag and Au, where θ is equal to $(T-T_m)/T_m$. All superclusters containing magic atom numbers have their molar volume obeying these laws at their critical formation temperature. The molar volumes $V_m$ for θ = 0.198 are maximum and equal to $13.4 \times 10^{-6}$, $19.68 \times 10^{-6}$, and $19.68 \times 10^{-6}$ m³/mole, for Cu, Ag and Au respectively.

They correspond to an infinite radius for superclusters in the absence of crystallization [15]. The molar volume $V_m$ of bulk superclusters would be attained when $\theta-\varepsilon_{nm}$ becomes equal to zero using the critical radius as a hidden variable becoming infinite instead of the temperature.

The Fermi energy change $\Delta E_F$ depends on $\Delta z$ in (32); $\Delta z$ is calculated with (24–26) for each radius $R = R^*_{2ls}(T)$ in (6), determining n from (28). Equation (33) is now respected for Cu, Ag and Au, as shown in Figure 10. The Fermi energy $E_F$ is defined in (34), assuming that there is one conduction electron per atom in Cu, Ag and Au:

$$E_F = \frac{\hbar^2}{2m_0}\left(\frac{3\pi^2}{V_m}\right)^{2/3}, \qquad (34)$$

where $V_m$ is the liquid molar volume at $T_m$ which is equal to $7.95 \times 10^{-6}$, $11.5 \times 10^{-6}$, and $11.3 \times 10^{-6}$ m$^3$/mole for Cu, Ag and Au respectively [64].

**Figure 10. Relative Fermi energy change between superclusters and liquid**. The $\Delta E_F/E_F$ is plotted as a function of the relative volume change $\Delta V_m/V_m$ of superclusters, where $V_m$ is the molar volume of the supercluster having an infinite radius.

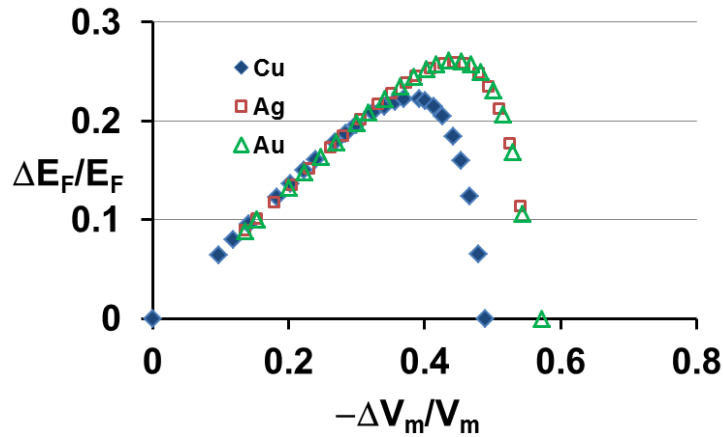

The molar electronic specific heat $C_{el} = \gamma_{el} \times T$ of Cu, Ag and Au superclusters can be obtained from the knowledge of their electronic density of states $D(E_F)$ at the Fermi level, calculated with (35) and (36) [64]:

$$\frac{\Delta E_F}{m} m D(E_F) = N_A \Delta z \qquad (35)$$

$$\gamma_{el} = \frac{\pi^2}{3} D(E_F) k_B^2 \qquad (36)$$

The $\Delta z$ values have been previously determined from (24–26). Each n-atom supercluster has its own molar volume $V_m$ and its own $\Delta z$ at $T_m$ is determined with (24) = (25) with $R = R^*_{2ls}(T_m)$ depending on $V_m$ and $\varepsilon_{nm0} = 0.217$. The electronic specific heat coefficient $\gamma_{el}$ is plotted in Figure 11 as a function of the supercluster molar volume.

The electronic specific heat coefficient $\gamma_{el}$ of superclusters falls when their molar volume $V_m$ and their radius decrease below $T_m$. The coefficients $\gamma_{el}$ of Cu, Ag and Au crystals at 4 K are a little larger, being equal to 0.695, 0.646 and 0.729 instead of 0.48, 0.547 and 0.599 mJ/K$^2$/mole at $T_m$ respectively [64]. Small crystals are known to become insulating for radii smaller than 5 nm when they are studied out of their melt [10]. This electronic transformation is also present in superclusters and is very abrupt below their critical growth volume at $T_m$ as shown in Figure 11. The $\gamma_{el}$ at $T_m$ is also calculated as a function of the supercluster radius R and represented in Figure 12. The coefficient $\Delta z$ is obtained at $T_m$ with (24) = (25) and $\varepsilon_{nm} = \varepsilon_{ls0} = 0.217$. Then, the potential energy $U_0$ depending on R is known for each value of R and the quantified coefficient $\varepsilon_{nm0}$ of an n-atom supercluster of radius R is deduced from (25,26). In Figure 12, the highest points are calculated at $T_m$ while the lowest are already shown in Figure 11. The smallest superclusters are still metallic at $T_m$, while they become insulating when the temperature is close to $T_m/3$. All these predictions are in good agreement with many properties of divided metals. They are only based on an enthalpy saving equal to $0.217 \times (1 - 2.25 \times \theta^2) \times \Delta H_m$ for the supercluster formation in all liquid elements.

**Figure 11. Electronic specific heat of superclusters.** The supercluster electronic specific heat coefficient in mJ/K$^2$/mole is plotted versus their molar volume in m$^3$ when the critical radius is smaller and smaller below $T_m$.

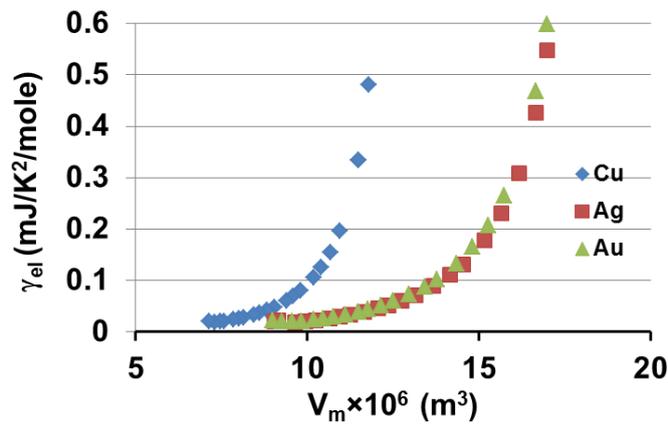

**Figure 12. Electronic specific heat coefficient of Cu, Ag, and Au superclusters** as a function of supercluster radius R at $T = T_m$ (colored points) and for $T < T_m$ when R is the critical radius (black points).

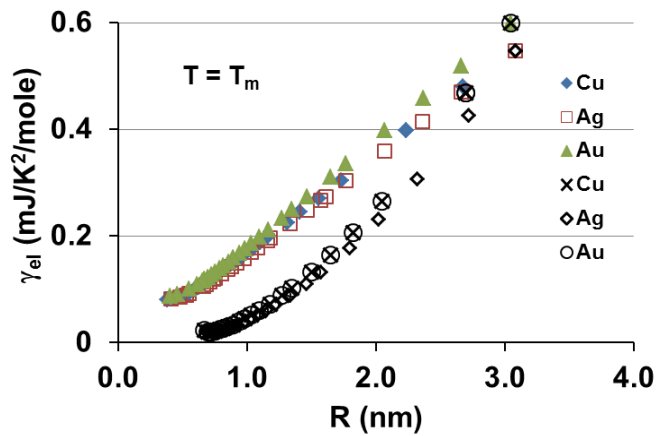

**8. Silver supercluster formation into and out of undercooled liquid**

The formation of icosahedral silver clusters with magic numbers n of atoms equal to 13, 55, 147, 309, 561, 923, 1415 and 2057 has been already studied out of liquid by molecular dynamics in the temperature range 0–1300 K. Icosahedral clusters of 13, 55 and 147 are formed below room temperature and larger clusters with n= 309, 561, 923, 1415 are formed from 300 to 1000 K. The radii of these Ag stable superclusters have been found to be equal to 2.74, 5.51, 8.32, 11.14 and 14.94 Å for n = 13, 55, 147, 309, 561 respectively [5]. The Ag radii have also been calculated in the liquid using the molar volume shown in Figure 9 and their formation temperature as deduced from the critical radius. Their values for n = 13, 55, 147, 309, 561 are nearly equal to those predicted by molecular dynamics, as shown in Table 4 and Figure 13.

**Table 4. The Ag supercluster radii with magic atom numbers**. The radius R is deduced from molar volume $V_m$ and equal to critical radius $R^*_{2ls}(T)$ given in (6). For $T > T_m/3 = 411.33$ K, the energy saving coefficient $\varepsilon_{ls}$ in (17) is used with $\varepsilon_{ls0} = 0.217$ and $(\theta_{0m})^{-2} = 2.25$. For $T < 411.33$ K, $\varepsilon_{ls}$ is equal to zero. The radii $R_{MD}$ result from molecular dynamics simulations [5].

| n | 13 | 55 | 147 | 309 | 561 | 923 | 1415 | 2057 |
|---|---|---|---|---|---|---|---|---|
| R (Å) | 3.387 | 5.485 | 8.541 | 11.63 | 14.67 | 17.68 | 20.67 | 23.66 |
| $R_{MD}$ (Å) | 2.74 | 5.51 | 8.32 | 11.14 | 14.94 | | | |
| T (K) | 0 | 291 | 604 | 817 | 952.2 | 1043.5 | 1108.4 | 1234 |

**Figure 13. The critical atom number in blue versus the critical radius and** $R_{MD}$ **the radius** calculated by molecular dynamics simulations in red square [5].

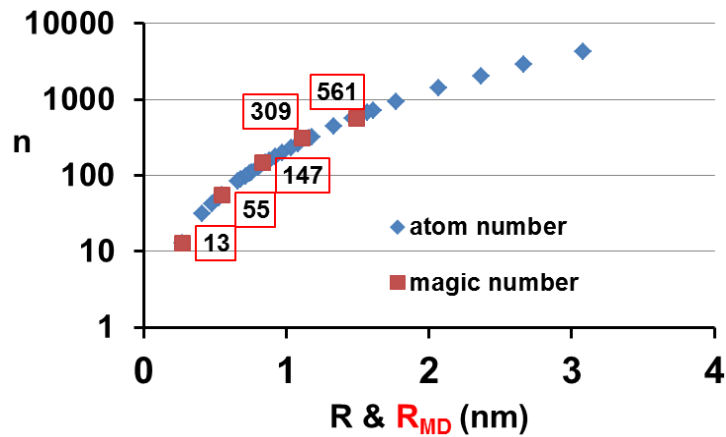

## 9. Melting of Cu, Ag and Au superclusters varying the superheating times  *9.1.*

*9.1 Overheating of Cu, Ag and Au superclusters*

The melting temperatures are now calculated using the molar volume associated with the supercluster radius as shown in Figure 9. The superheating time continues to be equal to 600 seconds. The supercluster radius variation is continuous while the radius of magic number clusters is indicated in Figures 14, 15 and 16. In these three figures, the Cu, Ag, and Au supercluster radius is plotted versus the reduced temperature $\theta = (T-T_m)/T_m$. The points labeled "homogeneous" are calculated assuming that supercluster melting is produced by liquid homogeneous nucleation using (29,30). The triangles labeled (n-13) are calculated assuming that the supercluster melting is induced by previous formation of liquid droplets of 13 atoms into superclusters using (31). The homogeneous nucleation temperatures are much too high compared to the (n-13) temperatures. The undercooling temperatures depend on the volume sample v. The square points are determined for $\ln(K_{ls}\cdot v\cdot t_{sn}) = 71.8$ corresponding to $v\cdot t_{sn} = 12\times 10^{-9}$ m$^3$.s and a heterogeneous nucleation induced by superclusters of radius R when the applied superheating temperature is smaller than those indicated by triangles. Another supercooling temperature represented by triangle points is added in Figures 15 and 16. In Figure 15, $v\cdot t_{sn}$ is equal to $7.08\times 10^{-22}$ m$^3$.s while, in Figure 16, $v\cdot t_{sn} = 15\times 10^{-7}$ m$^3$.s. These three figures show that an undercooling rate of about 20% is generally obtained when the sample volume is of the order of a few mm$^3$ and the applied superheating rate is less than about 25%. The undercooling temperature is very stable when the superheating is less than 25%. Larger undercooling rates are obtained using much smaller volume samples [17].

**Figure 14. Supercooling temperatures of liquid copper** controlled by unmelted superclusters having melting temperatures depending on overheating rate applied during 600 s.

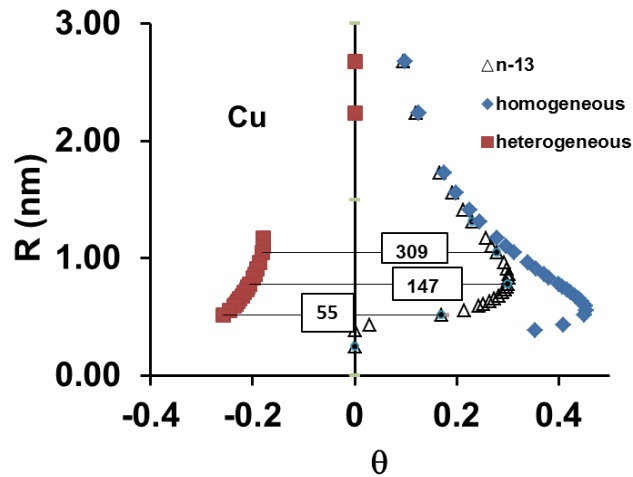

**Figure 15. Supercooling temperatures of liquid** silver depending on sample volume and controlled by unmelted superclusters having melting temperatures depending on overheating rate applied during 600 s.

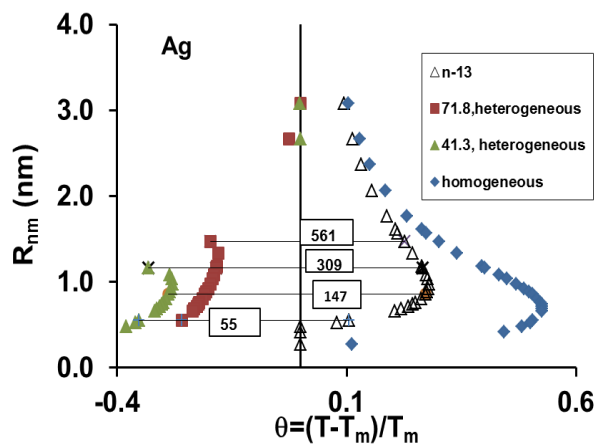

**Figure 16. Supercooling temperatures of liquid gold** depending on sample volume and controlled by unmelted superclusters having melting temperatures depending on overheating rate applied during 600 s.

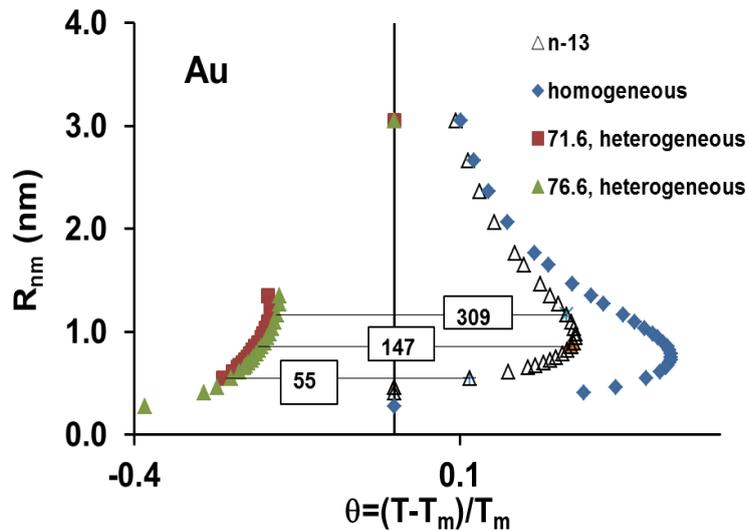

*9.2. Analysis of the influence of Cu superheating time on the undercooling rate*

The superheating time has a strong influence on supercluster melting, as shown by studies of Cu undercooling [51]. It has been found that a minimum superheating temperature of 40 K is required in order to achieve any undercooling prior to crystallization nucleation. This phenomenon is also observed in many magnetic texturing experiments [65]. In Table 3, the first Cu supercluster to be melted at $\theta = 0.033$ in 600 s, corresponding to a superheating of 44.7 K, contains 13 atoms in perfect agreement with the observation. There is no other supercluster melting. A temperature below which no small supercluster melts is predicted in this model. The lowest value of the undercooling temperature is obtained when 6 thermal cycles are applied prior to nucleation after 6 steps of 2400 s at 1473 K. The total time evolved at 1473 K is 14400 s. In Figure 17, the time necessary to melt all superclusters surviving in copper melt is calculated. With $\ln K_{ls} = 89.26$ instead of 90, the time to melt all the 13-atom clusters is 141 s, while that to melt 55-atom clusters is 14541 s, which is in very good agreement with the measurements [51]. The other superclusters are melted in very short times after the melting of the 55-atom clusters. The reduced undercooling temperature becomes equal to $\theta = -0.259$ after these thermal treatments of a sample of 5.7 mm in diameter. A homogeneous nucleation of 13- and 55-atom clusters leads to $\theta = -0.252$. These results show that superclusters can be chain melted with a weaker superheating if the time evolved at the overheating temperature is increased substantially beyond 600 s. Our model can be used to evaluate the approximate superheating time leading to a thermodynamic equilibrium of a melt without condensed superclusters at any temperature above $T_m$.

**Figure 17. Chain melting of superclusters versus superheating time**. The first cluster to be melted at θ = 0.086 contains 13 atoms, the next ones 55, 147, 309 and 561 because the liquid droplet grows in the core of the largest particles.

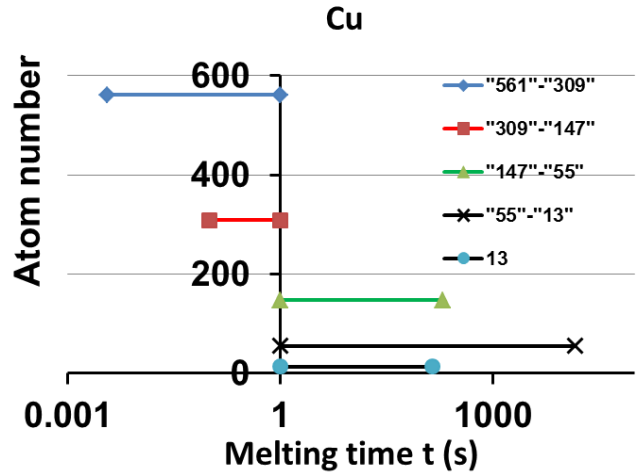

## 10. Conclusions

The undercooling temperatures of 32 of the 38 liquid elements are predicted for the first time in good agreement with experimental values depending on the sample volume, without using any adjustable barrier energy, and only assuming the existence of growth nuclei containing stable magic atom numbers n equal to 13, 55, 147, 309 and 561 that are generally devoted to icosahedral structures. The model is based on a volume enthalpy saving $\varepsilon_v$ previously determined to be equal to $0.217 \times \Delta H_m/V_m$ at $T_m$ and added to the classical Gibbs free energy change for a critical nucleus formation in a melt. This enthalpy is due to the Laplace pressure change $\Delta p$ acting on the growth nuclei and equalizing the Fermi energies of liquid and nuclei in metallic liquids. The Gibbs free energy change has to contain a contribution $-V_m \times \Delta p$ which has been neglected up to now because its magnitude was unknown. This missing enthalpy has serious consequences because the critical radius for crystal growth is considered, in the classical view, as being infinite at the melting temperature and all solid traces being eliminated in melts. This is in contradiction with many experiments on the superheating influence on undercooling rates and on magnetic texturing efficiency [51,65–69]. Nuclei having radii smaller than the critical radius at $T_m$ are melted at higher temperatures depending on the superheating time and on their atom number.

Some growth nuclei survive above $T_m$ because they are superclusters that are not melted by surface melting. This new property of superclusters is a consequence of the thermal variation of $\varepsilon_v$, which is a unique function of $\theta^2 = [(T-T_m)/T_m]^2$ being maximum at $T_m$, and a fusion heat equal to that of bulk crystals. The surface atom fusion heat is not weakened and there is no premelting of these entities depending on their radius. This thermal variation was established, for the first time, from our study of the maximum undercooling rate of the same liquid elements. In addition, it is the only law validating the existence of non-melted intrinsic entities.

The energy saving is proportional to the supercluster reverse radius $R^{-1}$ when n ≥ 147 and is quantified for n < 147. The quantified energy at $T_m$ is calculated by creating a virtual s-electron transfer from the nucleus of radius R to the melt and an electrostatic spherical potential induced by the surface charges and also varying with $R^{-1}$. The Schrödinger equation solutions are known and used to predict the condensation temperatures of 13-atom superclusters in undercooled melts which govern the crystallization temperatures of liquids having fusion entropy larger than 20 J/K/mole.

The superclusters are melted by homogeneous or heterogeneous liquid nucleation in their core. The liquid homogeneous nucleation is effective in all superclusters when $\Delta S_m \geq 20$ J/K/mole while a chain melting is produced, starting with a 13-atom droplet induced in the core of the supercluster and being magnified with the time increase at the superheating temperature. The model is able to predict an approximate value of the minimum time necessary to melt superclusters and to attain the true thermodynamic equilibrium of the melt at any superheating temperature.

The electronic specific heat of superclusters submitted to Laplace pressure in metals is determined for the first time from the enthalpy saving deduced from undercooling experiments. It strongly declines with radius as compared to that of a bulk metal, in agreement with the conductance properties of tiny clusters having radii smaller than 5 Å. The electronic s-state density of superclusters is greatly weakened compared to that of bulk crystals when their radius decreases. The supercluster critical radii deduced from the nucleation model are in quantitative agreement with recent molecular dynamics simulations devoted to Ag cluster radii.

The transformation of superclusters in crystals occurs for a radius between the critical radius for crystal growth and that for supercluster growth because the superclusters have a much lower density than crystals. The Gibbs free energy change from the liquid state to crystal becomes smaller than that of the supercluster just above its maximum at the crystal critical radius.

**Acknowledgments**:

Thanks are due to Professors Bernard Dreyfus and Jacques Friedel. The idea of this work started 50 years ago with criticisms of Bernard Dreyfus who considered that the nucleation equation would have to depend on the Fermi energy of conduction electrons. My first model developed in 2006, introduces an energy saving in the classical Gibbs free energy change for a nucleus formation that I had attributed to a transfer of conduction electrons from nucleus to liquid. Jacques Friedel noted that this transfer does not exist. I have proposed in 2011 that a Laplace pressure change acting on growth nuclei induces the enthalpy saving and the equalization of Fermi energies without electron transfer. A virtual transfer is nethertheless considered to quantify the energy saving in agreement with undercooling rates. The author thanks Dr Andrew Mullis for the confirmation that the copper sample weights used in [51] are equal to 0.6 to 1 g.